\documentclass[%
 reprint,
 prd,
 showpacs,
 amsmath,amssymb,
 aps,
 lengthcheck,
floatfix,
]{revtex4-1}
\usepackage{bibunits}

\usepackage{graphicx}
\usepackage{dcolumn}
\usepackage{bm}
\usepackage{hyperref}
\usepackage[mathlines]{lineno}

\usepackage{soul}

\usepackage{graphicx}
\usepackage{times}
\usepackage[normalem]{ulem}
\usepackage{color}
\usepackage{cellspace}
\usepackage{multirow}

\newcommand{\comment}[1]{}
\renewcommand\sout{\bgroup \color{red} \ULdepth=-.5ex \ULset}

\usepackage{rotating}
\usepackage{adjustbox}
\usepackage{makecell}
\usepackage{xcolor}
\usepackage{tikz,xcolor,hyperref}

\definecolor{lime}{HTML}{A6CE39}
\DeclareRobustCommand{\orcidicon}{%
	\begin{tikzpicture}
	\draw[lime, fill=lime] (0,0) 
	circle [radius=0.16] 
	node[white] {{\fontfamily{qag}\selectfont \tiny ID}};
	\draw[white, fill=white] (-0.0625,0.095) 
	circle [radius=0.007];
	\end{tikzpicture}
	\hspace{-2mm}
}

\def\simge{\mathrel{\rlap{\raise 0.511ex
     \hbox{$>$}}{\lower 0.511ex \hbox{$\sim$}}}}
\def\simle{\mathrel{\rlap{\raise 0.511ex
      \hbox{$<$}}{\lower 0.511ex \hbox{$\sim$}}}}

\usepackage{braket}

\newcommand{\pset}{\vartheta}
\newcommand{\likelihood}{p(\vec{d}(t)|\vec{\pset},H)}
\newcommand{\prior}{p(\vec{\pset}|H)}
\newcommand{\evidence}{p(\vec{d}(t)|H)}
\newcommand{\posterior}{p(\vec{\pset}|\vec{d}(t),H)}
\newcommand{\fpeak}{f_{\mathrm{peak}}}

\begin{document}

\title{Inferring physical properties of stellar collapse by third-generation gravitational-wave detectors}

\author{Chaitanya~Afle$^{1,2}$\href{https://orcid.org/0000-0003-2900-1333}{\orcidicon}}
\email{chafle@syr.edu}
\author{Duncan~A.~Brown$^{1,2}$\href{https://orcid.org/0000-0002-9180-5765}{\orcidicon}}

\affiliation{$^1$ Department of Physics, Syracuse University, Syracuse, NY 13244, USA}
\affiliation{$^2$ Kavli Institute for Theoretical Physics, University of California, Santa Barbara, CA 93106, USA}
\begin{abstract}
Galactic core-collapse supernovae are among the possible sources of gravitational waves. We investigate the ability of gravitational-wave observatories to extract the properties of the collapsing progenitor from the gravitational waves radiated. We use simulations of supernovae that explore a variety of progenitor core rotation rates and nuclear equations of state and examine the ability of current and future observatories to determine these properties using gravitational-wave parameter estimation. We use principal component analysis of the simulation catalog to determine the dominant features of the waveforms and create a map between the measured properties of the waveform and the physical properties of the progenitor star. We use Bayesian parameter inference and the parameter map to calculate posterior probabilities for the physical properties given a gravitational-wave observation. We demonstrate our method on a random sample of the waveform catalog that was excluded from construction of the principal component analysis and estimate the ratio of the progenitor's core rotational kinetic energy to potential energy ($\beta$) and the post bounce oscillation frequency. For a supernovae at the distance of the galactic center (8.1 kpc) with $\beta = 0.02$ our method can estimate $\beta$ with a $90\%$ credible interval of $0.004$ for Advanced LIGO, improving to $0.0008$ for Cosmic Explorer, the proposed third-generation detector. We demonstrate that if the core is rotating sufficiently rapidly for a signal source within the Milky Way observed by Cosmic Explorer, our method can also extract the post bounce oscillation frequency of the protoneutron star to a precision of within $5$~Hz ($90\%$ credible interval) allowing us to constrain the nuclear equation of state. For a supernovae at the distance of the Magellanic Clouds (48.5 kpc) Cosmic Explorer's ability to measure these parameters decreases slightly to $0.003$ for rotation and $~11$~Hz for the postbounce oscillation frequency ($90\%$ credible interval). Sources in Magellanic Clouds with $\beta < 0.02$ will be too distant for Advanced LIGO to measure these properties.
\end{abstract}

\maketitle


\section{Introduction}
When the core of a massive star exceeds its Chandrasekhar mass, it begins to undergo gravitational collapse \cite{Bethe:1990mw, Colgate:1966ax, Woosley:2002zz, Heger:2002by}. The core-collapse and subsequent bounce can power a supernovae explosion that radiates light, neutrinos, and gravitational waves (see e.g. Refs. \cite{Janka:2012wk,Muller:2019upo,Fryer:2011zz,Burrows:2012ew} and references therein). Gravitational waves generated during the supernovae travel unhindered through the stellar envelope, carrying information about the structure and dynamics of the collapsing star. Advanced LIGO will be able to detect core-collapse supernovae out to 50 kpc if the cores are rapidly rotating and the explosion is magnetorotationally driven, and to 5 kpc if the explosion is neutrino driven \cite{Gossan:2015xda, Abbott:2019pxc}. Cosmic Explorer, a proposed third-generation detector will be able to observe neutrino driven explosion signals out to a few hundred kiloparsecs \cite{Srivastava:2019fcb}, and the magnetorotationally driven explosion signals out to 2~Mpc. The estimated event rate for core-collapse supernovae in the Milky Way is 1-3 per century \cite{Bergh:1991ek, Cappellaro:1996cc, Li:2010kd, Diehl:2006cf}. While the probability of observing a signal within the reach of these detectors is low, if the information about the supernova can be extracted from the gravitational waves, it would shed new light on the physical processes of core-collapse.

Significant advances have been made over the last two decades in the simulation of core-collapse supernovae (see e.g. Refs. \cite{Janka:2016fox, Burrows:2019zce} and references therein). Abdikamalov \textit{et al.}~\cite{Abdikamalov:2013sta} performed 132 simulations in which they studied the dependence of the gravitational-wave signal at the core bounce and postbounce on the rotational properties of the progenitor core. They quantify rotation of the core by the ratio of the rotational kinetic energy and the gravitational potential energy $\beta = T/|W|$ and find that the gravitational-wave strain amplitude at the bounce primarily depends on $\beta$, while the degree of differential rotation only becomes relevant for cores with $\beta \gtrsim 0.08$. They use two equations of state (LS220 and HShen) and explore the difference between the waveforms associated with the two equations of state. Richers \textit{\textit{et al.}}~\cite{Richers:2017joj} used the progenitor star identical to Ref.~\cite{Abdikamalov:2013sta} in their simulations. They investigated the dependence of the gravitational-wave signal on the nuclear equation of state. They performed a total of 1764 simulations exploring 18 equations of state and 98 rotation profiles (varying $\beta$ and differential rotation). They confirm that the gravitational-wave signal at the bounce is most sensitive to $\beta$, while the postbounce oscillations depends on the equation of state, which manifests itself through the characteristic frequency of the oscillations, $\fpeak$.

Abdikamalov \textit{et al.} attempted to determine if gravitational-wave observations could be used to extract physical information about the core rotation. They constructed a template bank of waveforms spanning the range of rotation rates in their simulations, projected signals against this bank, and found that a signal observed at 10~kpc by Advanced LIGO could be used to  constrain $\beta$ to within $20 \%$ when $\beta \gtrsim 0.05$. Heng introduced the idea of using principal component analysis to model a set of supernovae waveforms, rather than using the waveforms themselves as a template bank \cite{Heng:2009zz}. Previous studies have used principal component analysis to infer the core-collapse explosion mechanism \cite{Logue:2012zw, Powell:2016wke, Powell:2017gbj, Roma:2019kcd}. 

Edwards \textit{et al.} \cite{Edwards:2014uya} used a principal component basis of the Abdikamalov \textit{et al.} waveform catalog and Bayesian parameter estimation \cite{Rover:2009ia} to determine if the core rotation $\beta$ could be extracted from the observation of a signal. Using a linear model, they fit the posterior means of the principal component coefficients to the known values of the physical parameter. Then they sample from the posterior predictive $t$-distribution to make probabilistic statements about $\beta$ estimation. They test their method on signals observed in Advanced LIGO with a signal-to-noise ratio of $20$ and are able to recover signals with $\beta = 0.02$ with $\beta = 0.05 \pm 0.03$, improving the accuracy of measurement to $\beta = 0.05 \pm 0.04$ for signals with $\beta = 0.05$, with average $90\%$ credible interval widths of $0.06$. 

In this paper, we use the waveform catalog of Richers \textit{et al.} to determine how accurately Advanced LIGO and the proposed third-generation detector Cosmic Explorer could extract information about the nuclear equation of state and the progenitor core rotation rate from observations of core-collapse supernovae. Since the progenitor cores of supernovae are expected to be rotating relatively slowly (core rotation periods $ \gtrsim 30$ s) \cite{Heger:2004qp, fuller2015spin, Ott:2005wh}, we focus on the waveforms in the Richers \textit{et al.} set with $0 \leq \beta < 0.07$. We use a total of 659 waveforms spanning 13 nuclear equations of state. We use principal component analysis to construct a model that captures the features of the Richers \textit{et al.} catalog and construct a map between the parameters measured by the principal component model and the physical parameters of the waveform $\fpeak$ and $\beta$. We use Monte Carlo methods to perform Bayesian parameter estimation to measure the posterior probability distribution of the principal component model parameters and the constructed map to transform these into the posterior probability distributions of the physical parameters. 

We find that for sources with $\beta \geq 0.02$ at a distance of $8$ kpc, $\beta$ can be estimated with a $90\%$ credible interval of $0.004$ for Advanced LIGO, and $0.0008$ for Cosmic Explorer detectors. The precision of measurement for signal sources at $48.5$ kpc observed in Cosmic Explorer deteriorates to $90\%$ credible interval of $0.003$. We can constrain $\fpeak$ for sources within the Milky Way galaxy to with $90\%$ credible interval of $5$~Hz for detections in the third-generation detectors, if the $\beta$ for the signal is more than $0.02$, thus allowing us to constrain the nuclear equation of state.  

This paper is organized as follows: In Sec.~\ref{section:pca} we describe the the construction of a principal component basis set using the Richers \textit{et al.} waveforms from which we withhold a random sample of 10\% to test our method. In Sec.~\ref{section:map} we describe the construction of the map between the parameters of the principal component model and the physical waveform. In Sec.~\ref{section:pe} we describe our Bayesian parameter estimation methods, and in Sec.~\ref{section:results} we present the results of the methods using simulated signals in Advanced LIGO and Cosmic Explorer. In Sec~\ref{section:conclusion} we summarize our findings and discuss directions for future work.

\section{Principal Component Analysis} 
\label{section:pca}
Principal component analysis extracts the dominant features from a set of waveforms as linearly independent principal components \cite{Heng:2009zz}. In this study, we use singular value decomposition to compute the principal components. A set of discretely and evenly sampled-in-time waveforms can be written as the columns of a matrix $D$ which can be written as
\begin{equation}                                                                                D = U \Sigma V^T, \label{eq:svd}
\end{equation}
where the matrices $U$ and $V$ contain the orthonormal eigenvectors of $DD^T$ and $D^T D$, respectively, and the diagonal matrix $\Sigma$ contains the eigenvalues of $DD^T$. The orthonormal vectors in the matrix $U$ are the principal components, and are sorted in decreasing order of the size of the square root of the eigenvalues. Hence, the first principal component describes the most dominant feature in the set of waveforms. If we have $N$ waveforms in the catalog $D$, then $U$ contains $N$ principal components. By constructing a principal component decomposition of the catalog, we attempt to construct a set of basis vectors that captures the features of signals that lie in the space spanned by the waveform catalog, without requiring modelling every possible core-collapse in the catalog space. The principal component analysis provides us with a semianalytic model for core-collapse waveforms, given by
\begin{equation}                                                                                
H \approx \sum_{j=1}^{N} \alpha_{j} U_{j},
\end{equation}
where the $\alpha_{j}$ are the coefficients of the signal $H$ expressed in terms of the basis vectors $U_j$. We can use Bayesian parameter estimation to construct posterior probability densities on the model parameters $\alpha_{j}$ and hence the gravitational-wave signal $H$. However, there are two challenges to directly implementing this approach. First, the number of waveforms used to construct the principal component analysis $N$ must be large enough to accurately explore the features in the catalog (typically of order $10^2$--$10^3$ waveforms), but this $N$ may be significantly larger than the number of basis vectors needed to capture the essential features of the waveforms. Second, the measured $\alpha_{j}$ are parameters of the basis vectors and are not directly related to physical parameters of the waveforms. As suggested in previous works, we address these challenges in two ways. Since the principal component analysis tells us which basis vectors capture the dominant features of the catalog, we can construct an approximation to each waveform $h$ as a linear combination of a subset of the principal components
\begin{equation}                                                                                
h = \sum_{j=1}^{k} \alpha_{j} U_{j}. \label{eq:waveform_from_pcs}
\end{equation}
where $k < N$. Here, we use two approaches to choose the value of $k$; we study the overlap between the original waveforms in the catalog and approximations to these waveforms using a subset of basis vectors. If the overlap is unity, then the approximate decomposition exactly reproduces the original waveforms. We use the overlap method to make an initial choice of the number of basis vectors $k$ and then perform parameter estimation to confirm that the choice is sufficient; that is statistical error dominates over the systematic error that arises from choosing $k < N$. Finally, we determine which of the $\alpha_i$ are needed to extract the physical parameters $\beta$ and $\fpeak$ and use the catalog to construct the maps $\beta(\alpha_i)$ and $\fpeak(\alpha_i)$.

To construct the basis set, we use the axisymmetric general-relativistic hydrodynamic simulations from Richers \textit{et al.} that span 18 different equations of state and 98 rotation profiles \cite{Richers:2017joj}. They use a $12 M_{\odot}$ nonrotating progenitor (model s12WH07 from \cite{Woosley:2007as}) in the \texttt{CoCoNuT} code \cite{Dimmelmeier:2004me, Dimmelmeier:2002bm} once for each of the 18 equations of state. Richers \textit{et al.} imposed a rotation profile on the progenitor according to the cylindrical rotation law \cite{Zwerger:1997sq}:
\begin{equation}
    \Omega\left(r\right) = \Omega_0\left[1 + \left(\frac{r}{A}\right)^2\right]^{-1},
\end{equation}
where $A$ (measured in km) depicts the measure of degree of differential rotation, $\Omega_0$ is the maximum initial rotation rate, and $r$ is the distance from the rotational axis in km. 

We exclude the prompt convection part of the waveforms when building the principal component basis set. This part of the signal is highly stochastic in nature making it challenging to model with principal component analysis. However, the prompt convection phase is retained in the waveforms that are used as signals to test our method. Richers \textit{et al.} suggest that information on the progenitor core rotation and the equation of state can be extracted from the core bounce and the postbounce oscillations of the protoneutron star. We therefore use the criteria proposed by Richers \textit{et al.} to truncate the waveform $6$~ms after the third zero-crossing of the strain waveform after the bounce. We resample the waveforms to $16\,384$~Hz and ensure that the length of all waveforms is $1$~s by zero padding them with the core bounce aligned at $t=0.5$~s for all the waveforms. In our analysis, we only use the plus polarization of the waveforms.

\begin{figure}[t]
  \includegraphics[width=\columnwidth]{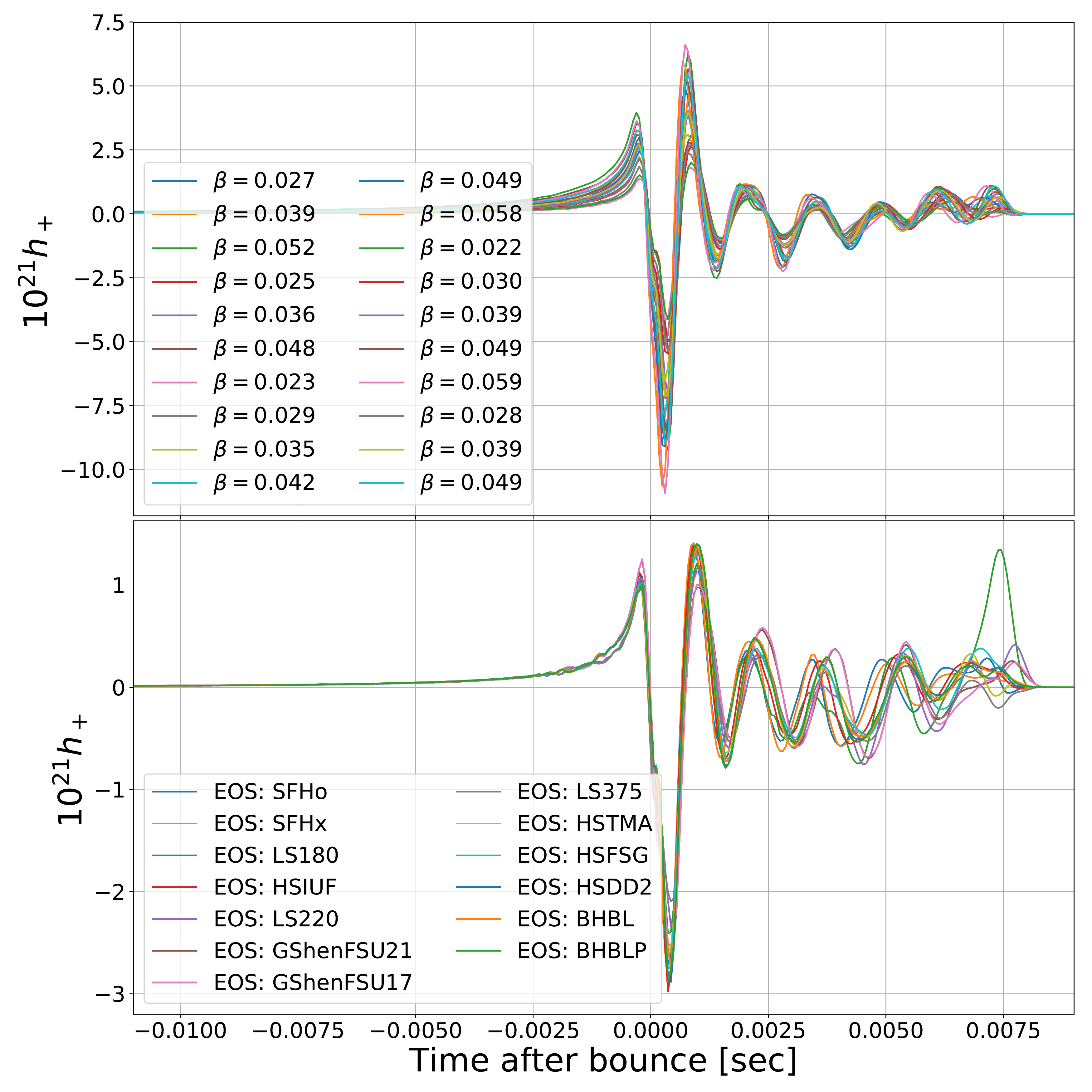}
  \caption{Gravitational wave strain assuming the distance to the progenitor of 10~kpc as function of time for bounce and postbounce oscillation phases of a core-collapse process. The waveforms are zero buffered to make them 1 second long, and the time of bounce is aligned at 0.5 seconds for all the waveforms. The top panel shows the waveforms for the \texttt{SFHx} equation of state with varying rotation rates between $\beta = 0.02$ and $\beta = 0.06$. The strain amplitude at the bounce increases with increasing $\beta$, while the postbounce oscillation frequency remains almost the same for all the waveforms corresponding to a given equation of state. Bottom panel shows the waveforms for $\Omega = 2.50 $ rad/sec and $A = 467$ km for the equations of state listed in Table \ref{tab:average_f_peak_values}. The bounce amplitude remains almost the same for the waveforms with the same core rotation rate, while the postbounce oscillation frequency varies for different equations of state.}
\label{fig:waveform_morphology}%
\vspace*{-0.5cm}%
\end{figure}

The general morphology of the waveforms can be seen in Fig.~\ref{fig:waveform_morphology}. Prior to the core bounce, the strain increases slowly. It decreases rapidly through the bounce to a local minimum. The depth of the local minimum increases with the rotation rate of the inner core at the time of the bounce. This phase is followed by the postbounce ringdown oscillations of the newly formed protoneutron star, which lasts $\sim 6$ ms. The characteristic frequency of these oscillations depends on the equation of state of the inner core. The top panel of Fig.~\ref{fig:waveform_morphology} shows the waveforms for \texttt{SFHx} equation of state and the rotation rates of the inner core between $\beta = 0.02$ and $0.06$. We can see that the depth of the first local minimum immediately after the core bounce increases with the rotation rate. However, the postbounce oscillations have almost the same frequency irrespective of the rotation rate. The bottom panel shows us the waveforms for $\Omega = 2.50$ rad/sec and the precollapse differential rotation rate $A = 467$ km for various equations of state listed in Table \ref{tab:average_f_peak_values}. We can note that the depth of the first local minimum is nearly the same for waveforms with different equation of state since the rotation rate is the same while the postbounce oscillation frequency is different for different equations of state. 

\begin{figure}[t]
  \includegraphics[width=\columnwidth]{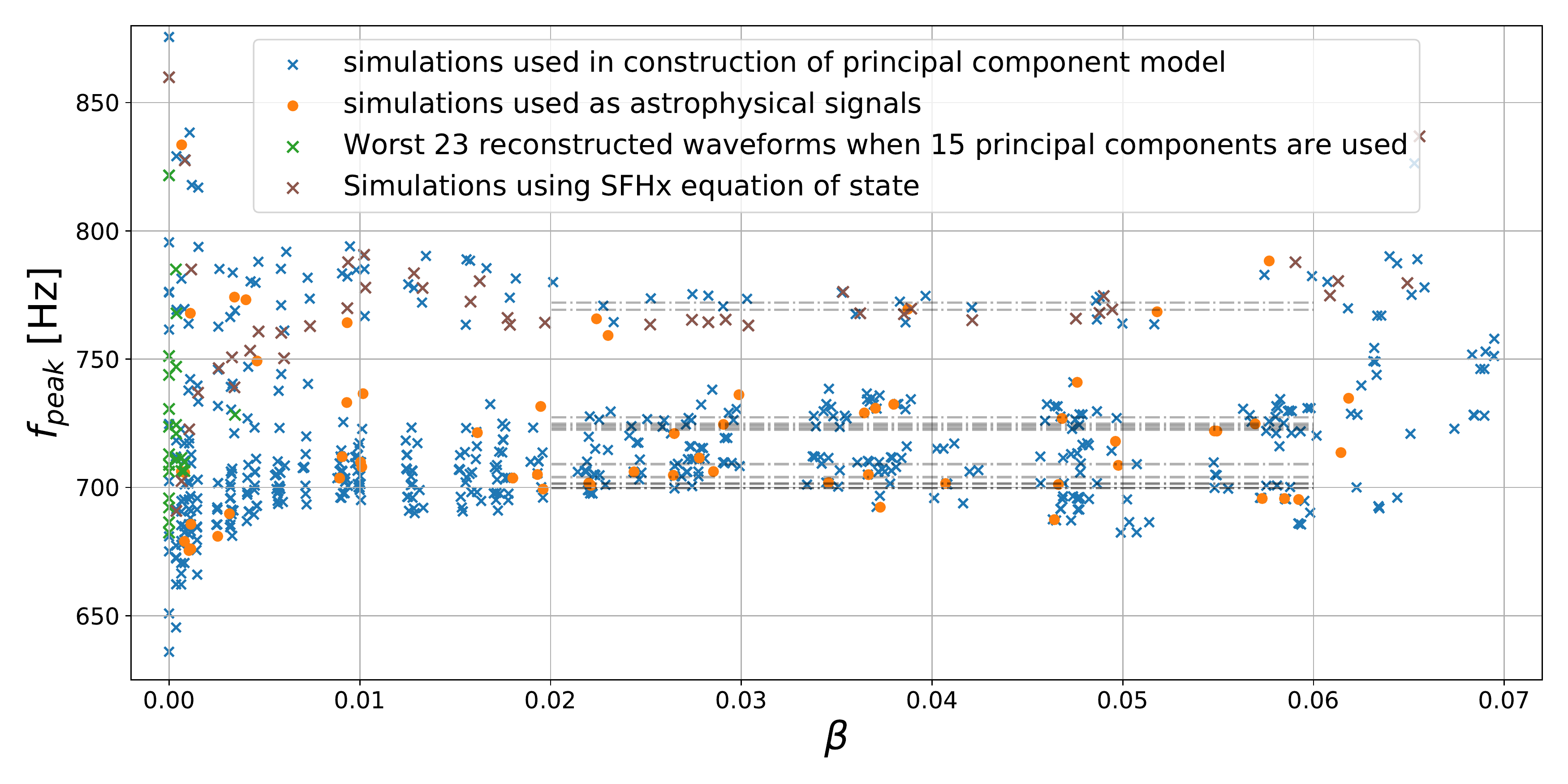}
  \vspace*{-0.5cm}%
  \caption{Frequency of postbounce oscillations is plotted on the vertical axis against $\beta$ of the waveforms on the horizontal axis. The crosses represent the waveforms that are used to build the principal component basis. This also includes the green crosses, showing the waveforms that are affected the most by only considering 15 principal components and not more. The simulations that use the \texttt{SFHx} equation of state are shown in brown crosses. The $\fpeak$ value for a given equation of state is independent of $\beta$ for $0.02 \leq \beta \leq 0.06$. The dashed lines represent the average  $\fpeak$ values of the waveforms of a given equation of state in this range, also given in Tab. \ref{tab:average_f_peak_values}.The orange dots represent the parameter values of the waveforms that are used as astrophysical signals in this study. }
\label{fig:f_peak_beta_with_worst_50}
\end{figure}

In order to focus on slowly rotating progenitor cores, we restrict the catalog to the set of simulations with $\beta<0.07$. We also exclude simulations whose equation of state is ruled out by observations of GW170817 \cite{TheLIGOScientific:2017qsa, De:2018zrk, Abbott:2018exr}, giving us $659$ waveforms in total. We select 60 waveforms at random from this set and reserve them for testing our methods; these test signals are not included in the construction of either the principal component decomposition or the map between principal component parameters and physical parameters.
We construct a principal component basis set from the remaining 599 waveforms. We do not consider the affects of the pre-collapse differential core rotation since Refs.~\cite{Abdikamalov:2013sta} and \cite{Richers:2017joj} show that the waveforms for slowly rotating cores are only very weakly dependent on the differential rotation profile. Therefore we consider parameterization of the catalog only by $\beta$, regardless of the differential rotation. 
Figure \ref{fig:f_peak_beta_with_worst_50} shows the values of $\beta$ and $\fpeak$ of the simulations used to construct the principal component analysis and map (crosses) and and the signals reserved to test our method (dots).

\begin{figure}[t]
  \includegraphics[width=\columnwidth]{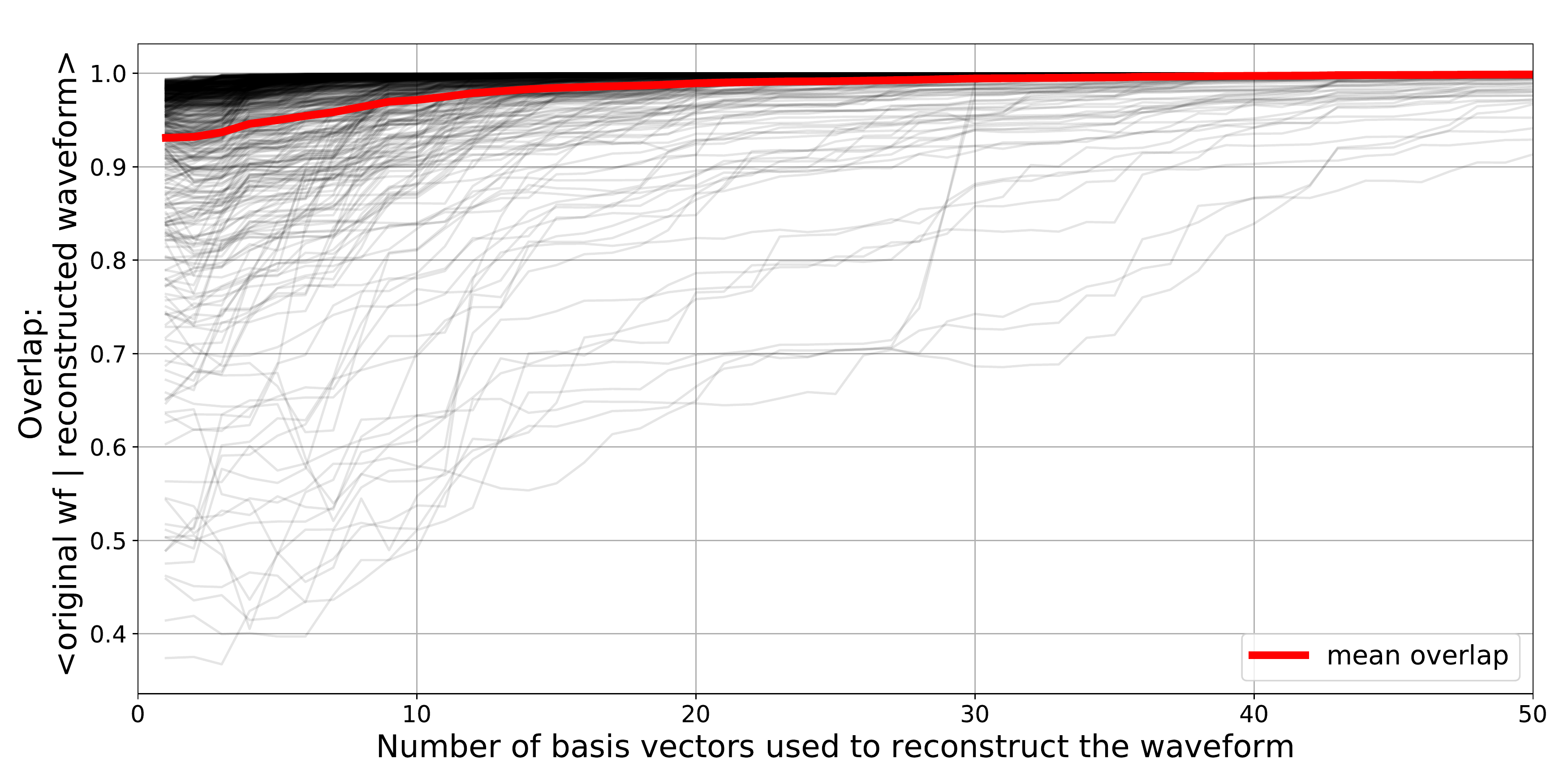}
  \caption{The plot shows how well can a given number of principal components (plotted on the horizontal axis) reconstruct the original waveform. We quantify this by computing the overlap between the original waveform and the reconstructed waveform, and show it on the vertical axis. Each of the waveforms is represented by a grey line, and the mean overlap of all the waveforms as a function of number of basis vectors used for construction is represented by the red line. 
\label{fig:sanity_check_pca}%
}
\end{figure}

Figure \ref{fig:sanity_check_pca} shows the reconstruction of each of the $599$ waveforms using the principal component basis set. The horizontal axis represents the number of principal components $k$ used to generate the waveform by Eq. \ref{eq:waveform_from_pcs} and the vertical axis represents the overlap between the original catalog waveform $H$ and the approximate reconstructed waveform $h$ for each value of $k$, where the overlap between is defined as \cite{Wainstein:1962}:
\begin{equation}
    \braket{H|h} = 4\mathbb{R} \int_0^{\infty} \frac{\tilde{h}(f)\tilde{H}(f)}{S_n (f)}\, \mathrm{d}x,
\end{equation} \label{eq:overlap}
where $\tilde{H}(f)$ and $\tilde{h}(f)$ are the Fourier transforms of the waveforms and $S_n (f)$ is the power spectral density of the Cosmic Explorer (CE1) detector noise. 
This figure shows that by using the first 50 of the 599 principal components, we are able to reconstruct the all 599 original waveforms with more than $90 \%$ overlap. 
However, we find that using $50$ basis vectors in the Bayesian parameter estimation is computationally expensive and note that if only $15$ basis vectors are used, $96\%$ of the waveforms are reconstructed with an overlap greater than $90\%$. In Fig.~\ref{fig:f_peak_beta_with_worst_50} the catalog waveforms for which 15 basis vectors are sufficient to reconstruct the overlap to $\geq 90\%$ are shown with blue crosses and the catalog waveforms that fail this criteria are shown with green crosses. We see that all the waveforms that require more than fifteen principal components to reproduce the waveform with at least $90\%$ overlap lie in the region of slowest core rotation $\beta$. These are the waveforms for which it is most challenging to extract $\beta$ and $\fpeak$ \cite{Richers:2017joj}. However, we still include these waveforms in our analysis.

Previous studies have used principal component analysis to construct a gravitational-waveform model for rotating core-collapse supernovae that is used for Bayesian reconstruction of the signal observed in the detector. R{\"o}ver \textit{et al.} \cite{Rover:2009ia} also used overlaps between the original waveforms and the waveforms generated through a subset of principal component basis to determine the number of basis vectors to be used in their waveform model. They used 128 waveform simulations from Dimmelmeier \textit{et al.} \cite{Dimmelmeier:2008iq} to construct their basis set and used 10 basis vectors. Edwards \textit{et al.} \cite{Edwards:2014uya} used a constrained optimization approach to select the number of basis vectors in their study. They used 132 waveforms in the Abdikamalov \textit{et al.} catalog \cite{Abdikamalov:2013sta} to construct their basis set and used the first 14 of the basis vectors in their model.

\section{Mapping to physical parameters} \label{section:map}
Having constructed a principal component model and determined that fifteen basis vectors are adequate to capture the essential features of the catalog space, we construct a map between the unphysical parameters of our model $\alpha_{j}$ and the physical parameters of interest $\beta$ and $\fpeak$. The ratio of the rotational kinetic energy to the gravitational potential energy of the inner core $\beta$, is a robust way of quantifying the rotation rate of the inner core \cite{Abdikamalov:2013sta, Richers:2017joj}. $\beta$ is a time dependent quantity that evolves during the core-collapse event. In our work we quantify the rotation rate of the core of the progenitor with $\beta$ at the time of the core bounce.  

\begin{figure}[t]
  \includegraphics[width=\columnwidth]{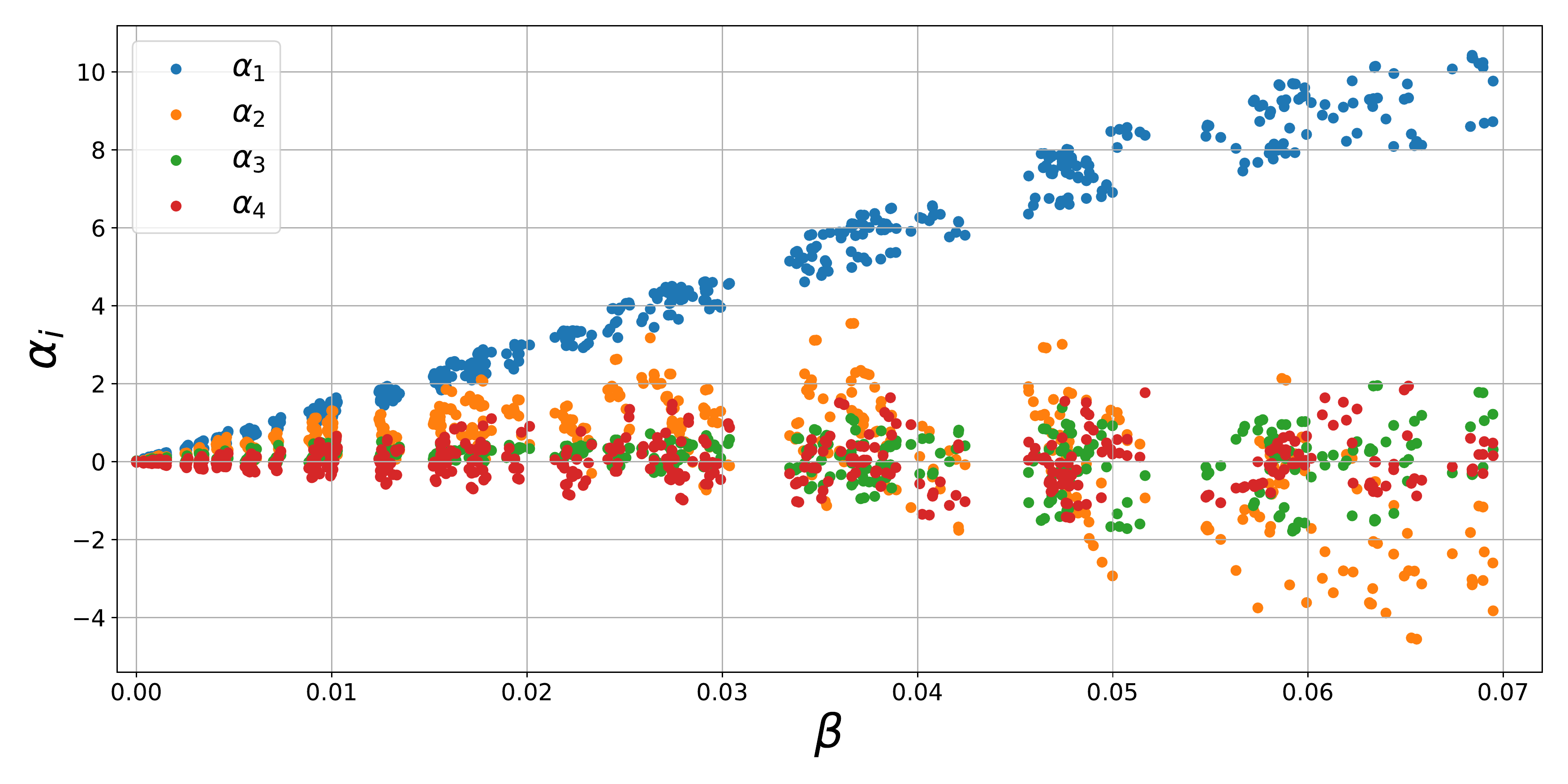}
  \caption{The coefficients of the first four principal components as a function of $\beta$. The coefficient of the first principle component, $\alpha_1$ (shown in blue) is most strongly correlated with $\beta$, exhibiting a roughly linear relation. The correlation between the other three coefficients and $\beta$ can be seen to be weaker. The values of the coefficients spread as $\beta$ increases because of different equations of state used in simulation of the waveforms. 
\label{fig:beta_alpha_map}%
\vspace*{-0.5cm}%
}
\end{figure}

Fig.~\ref{fig:beta_alpha_map} shows the values of the coefficients of the first four principal components $\alpha_i$ ($i = 1, 2, 3, 4$) as a function of the rotation rate $\beta$ for the waveforms in the catalog. We see that $\alpha_1$ is the parameter most strongly correlated with $\beta$, exhibiting a roughly linear dependence across the catalog space. The increase in the spread of points in $\alpha_1$ as $\beta$ increases is caused by waveforms with similar values of $\beta$ but different equations of state; the change in equation of state weakly affects the map between the two parameters. The correlation between the other three model parameters and $\beta$ is not as obvious. We use the data shown in Fig.~\ref{fig:beta_alpha_map} to construct a map $\beta(\alpha_1, \ldots, \alpha_k)$, where $k \leq 8$.

To construct the map using just the first model parameter $\beta(\alpha_1)$, we use the least square fit for a straight line, obtaining the slope $0.0326$ and the intercept $0.0007$. If we want to incorporate more than one model parameters to construct the map, we use interpolation to find $\beta(\mathbf{A)}$ for an arbitrary point $\mathbf{A} = (\alpha_1, \ldots, \alpha_n$) with $2 \leq n \leq 8$ using the known values of $\beta$ and $(\alpha_1, \ldots, \alpha_n)$. This interpolation is performed using the linear method of \texttt{scipy.interpolate.griddata} which finds the convex hull of $\mathbf{A}$, which consists of the nearest $n+1$ neighbours of $\mathbf{A}$ that contain $\mathbf{A}$: $\mathbf{A_1, \ldots, A_{n+1}}$, for which the $\beta$ values are known. $\mathbf{A}$ can be written as a weighted average of $\mathbf{A_1, \ldots, A_{n+1}}$:
\begin{equation}
 \mathbf{A} = \sum_{i=1}^{n+1} \gamma_i \mathbf{A_i},
\end{equation}
where $\gamma_i > 0$ and $\sum \gamma_i = 1$. The map for an arbitrary point is then generated using the linear interpolation with the $\gamma_i$s as the weights in the interpolation:
\begin{equation}
    \beta\left(\mathbf{A}\right)\approx\sum_{i=1}^{n+1}\gamma_{i}\beta\left(\mathbf{A}_{i}\right) .
\end{equation}
The interpolation fails if $\mathbf{A}$ does not lie within a convex hull of points with known values of $\beta$. Finding the convex hull of $\mathbf{A}$ becomes increasingly computationally expensive as the number of model parameters (and hence, the number of dimensions) used in the interpolation increases. To determine how many model parameters should be used in the map to construct a robust and sufficiently accurate map, we perform the following test. We first note that since our waveform catalog is large, the omission of one waveform from the construction of the principal component basis does not significantly change the principal component decomposition. Given this, we can exclude a waveform from the principal component analysis, construct the interpolation function using the remaining waveforms, and use this interpolating function to estimate the known value of $\beta$ for the waveform excluded from our algorithm. We repeat this procedure for each of the waveforms in the catalog used to construct the principal component basis and the interpolation function. Note that we do not use the 10\% of the catalog reserved for astrophysical testing here, as we reserve those waveforms for use until our method is fully tuned.

\begin{figure}[t]
  \includegraphics[width=\columnwidth]{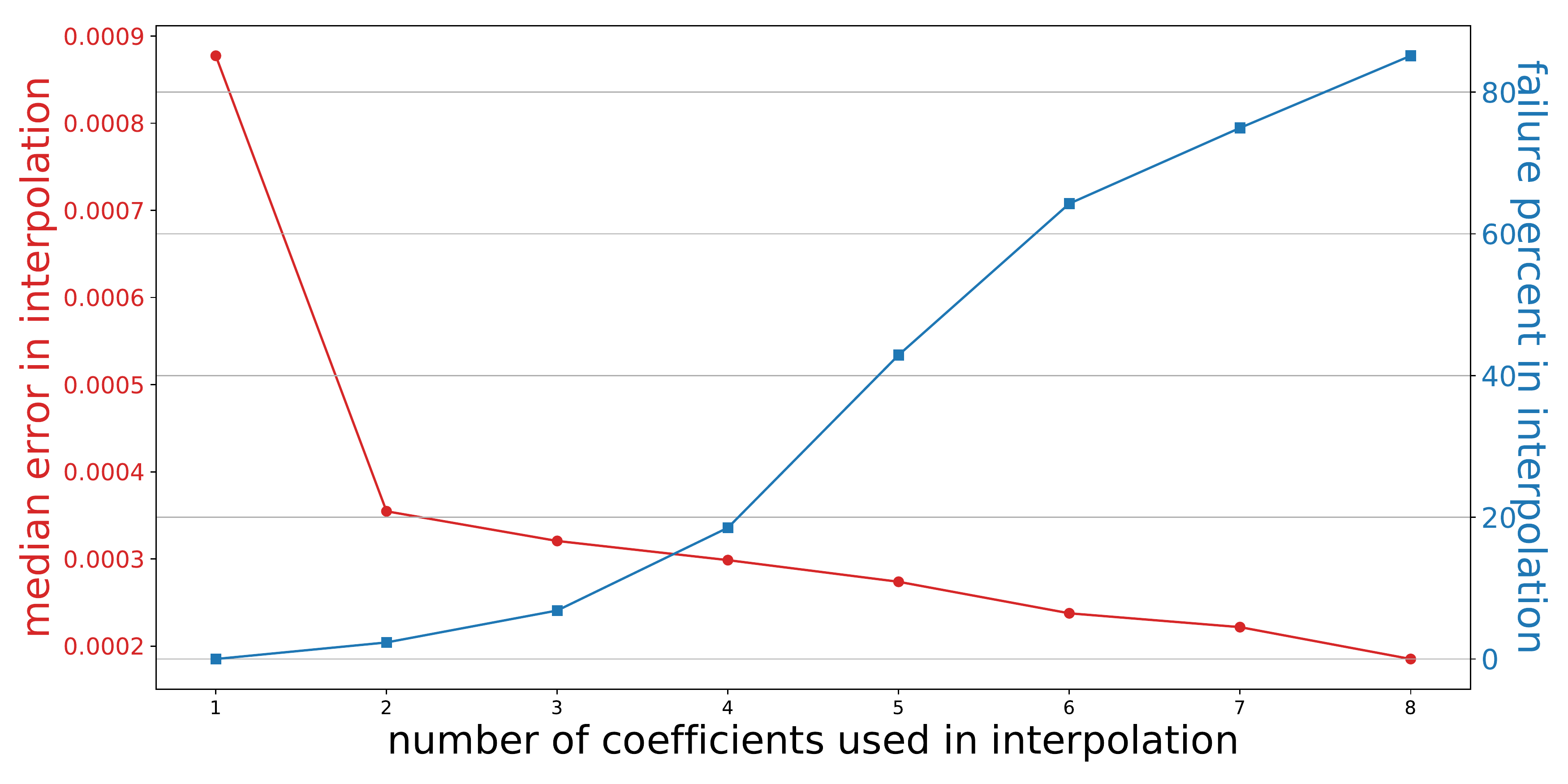}
  \caption{For each waveform in the catalog, a principal component basis set is constructed using all remaining waveforms. Using this basis set, $\beta(\alpha_1, \ldots, \alpha_k)$ maps are constructed using interpolation with the first $k=2, \ldots, 8$ model parameters, and $\beta$ of the excluded waveform is estimated using these maps. Least square fit for a straight line is used while using just the first model parameter to construct the map $\beta(\alpha_1)$. The median error in reconstructing $\beta$ through various maps and the respective failure rate in interpolation are plotted on the vertical axes. Using more number of model parameters reduces the error in interpolation, however increases the number of times the interpolation fails.
\label{fig:interpolation_check}%
\vspace*{-0.5cm}%
}
\end{figure}

The outcome of this test is shown in Fig.~\ref{fig:interpolation_check}. The horizontal axis shows the number of model parameters used to construct the map $\beta(\alpha_1, \ldots, \alpha_k)$ for $k \leq 8$. The median error in reconstructing $\beta$ from each of these maps for the waveforms in the catalog is plotted on the vertical axis. The failure rate of interpolation corresponding to each map is also shown. We see that as the number of model parameters used to construct the map increases, the interpolation error decreases. Maps that use interpolation with two or more model parameters have significantly less error as compared to the map $\beta(\alpha_1)$ constructed using the least square fit. Hence we do not use the map $\beta(\alpha_1)$ in our analysis. However, with increasing number of model parameters, the failure rate for interpolation also increases. The interpolation fails for more than $80\%$ of the cases when we use eight model parameters. The failure rate of the map constructed by using nine model parameters or more is even higher and we do not consider that in our analysis. We also note that the error in reconstruction of $\beta$ using the interpolation increases as $\beta$ increases. This can be attributed to the fact that that the volume of parameter space sampled is sparser as $\beta$ increases. 

\begin{figure}[t]
  \includegraphics[width=\columnwidth]{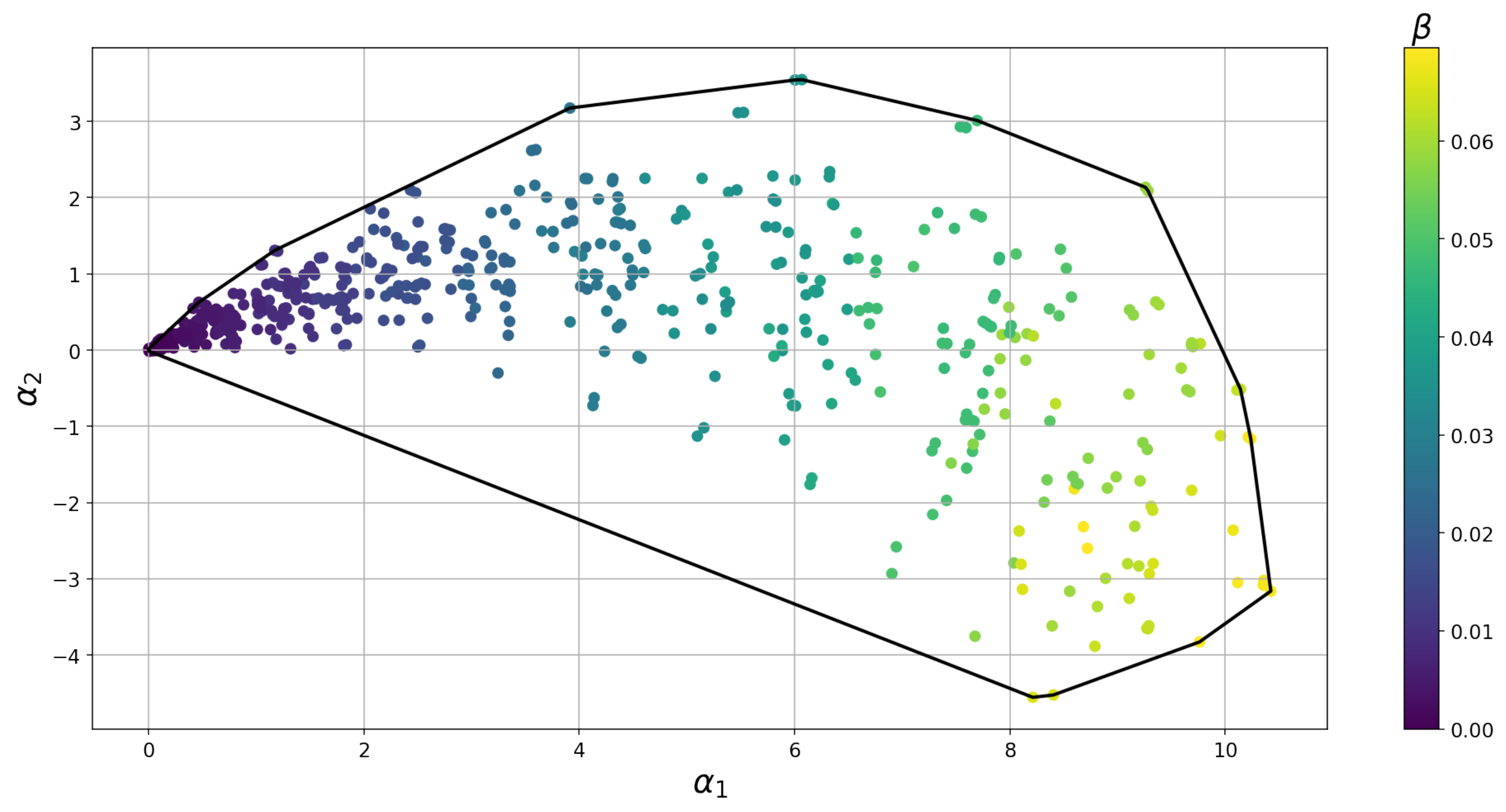}
  \caption{The $\alpha_2\ $ (vertical axis) vs $\alpha_1\ $ (horizontal axis) parameter plane for the waveforms in the catalog. The colorbar shows the $\beta$ corresponding to each of the waveforms. The two dimensional convex hull of the all the points is shown by the dashed black line. Interpolation fails for a point outside the convex hull. We can construct a three dimensional convex hull if we also incorporate $\alpha_3$. We constrain our MCMC samples to be within the three dimensional convex hull.
\label{fig:convex_hull}%
\vspace*{-0.5cm}%
}
\end{figure}

We use the maps $\beta(\alpha_1, ..., \alpha_k)$ with $k \leq 8$ to translate the posteriors obtained for the model parameters from the Bayesian inference of simulated signals to the posteriors on $\beta$. We constrain the samples to be in the convex hull of the first two model parameters, as shown in Fig.~\ref{fig:convex_hull} in order to successfully interpolate using the first three parameters. We first use the map constructed by using eight model parameters, which would result in some samples in the posteriors getting rejected because of the failure in interpolation. We then use the map formed by seven model parameters for the samples for which the interpolation failed previously, and repeat the procedure with maps constructed using fewer model parameters for the samples for which interpolation fails. Eventually, all the remaining samples are successfully interpolated by using the map $\beta(\alpha_1, \alpha_2, \alpha_3)$. Constraining the samples within the convex hull using four parameters or higher is computationally expensive. A much more robust map can be constructed by using machine learning and by populating the parameter space with more simulations. We leave the construction and testing of that map for future work. 

The postbounce oscillation frequency $\fpeak$ is the $l = 2$ f-mode peak frequency of the protoneutron star after the core bounce \cite{Ott:2012kr, Fuller:2015lpa}. Richers \textit{et al.} observed that for simulations with $0.02 \le \beta \le 0.06$, $\fpeak$ for a given nuclear equation of state is independent of the value of $\beta$ (see Fig.~\ref{fig:f_peak_beta_with_worst_50}), with the softer equations of state having a higher postbounce oscillation frequency. We use this relation between $\fpeak$ and the equation of state, shown in Table~\ref{tab:average_f_peak_values}, to infer the equation of state dependence on $\fpeak$. 
\begin{table}
\begin{center}
\begin{tabular}{c c c} 
\hline
\hline
Equation & $\fpeak$  & $\fpeak$ \\ 
of State & Mean value & Standard deviation \\
         & [Hz]      &  [Hz] \\
\hline   
SFHo        & 772.1 & 5.6  \\ 
SFHx        & 768.9 & 6.2  \\
LS180       & 728.4 & 6.4 \\
HSIUF       & 724.2 & 8.4  \\
LS220       & 723.7 & 6.4  \\
GShenFSU2.1 & 723.2 & 11.1 \\
GShenFSU1.7 & 721.1 & 10.3 \\
LS375       & 709.1 & 8.1  \\
HSTMA       & 704.1 & 5.7  \\
HSFSG       & 702.1 & 7.9  \\
HSDD2       & 701.6 & 8.3  \\
BHBLP       & 699.7 & 8.6  \\
BHBL        & 699.7 & 8.2  \\
\hline
\end{tabular}
\caption{The mean and standard deviation of the $\fpeak$ values of the waveforms used to form the principal component basis belonging to a particular equation of state with $0.02 \leq \beta \leq 0.06$.} \label{tab:average_f_peak_values}
\end{center}
\end{table}
To measure $\fpeak$, in our analysis, we the method of Richers \textit{et al.} We first isolate the postbounce oscillation from the earlier bounce and the later convection phases of the waveform by taking the Fourier transform of the waveform up to the end of the bounce phase $t_{\mathrm{be}}$ (taken to be the third zero crossing after the core bounce) and, separately, the Fourier transform of the waveform up to $t_{\mathrm{be}} + 6$ ms, in order to include a few cycles of the postbounce oscillations and isolate them from the convective phase. The Fourier transform of the waveform up to the bounce phase is subtracted from the Fourier transform that includes postbounce oscillations and the largest spectral feature within the window $600$ - $1075$ Hz is $\fpeak$. As found by Richers \textit{et al.}, for slowly rotating cores with $\beta \leq 0.02$ this method to extract $\fpeak$ is unreliable since the protoneutron star oscillations are only weakly excited. For $\beta \geq 0.06$, centrifugal forces start affecting the postbounce oscillations and the $\fpeak$ value depends on differential rotation in addition to the equation of state.

In our analysis, we measure $\fpeak$ of a signal observed in a detector by applying the method of Richers \textit{et al.} to the waveform reconstructed by our Bayesian parameter estimation. For each sample in our posterior probability distribution, we construct the approximate signal given by Eq.~\ref{eq:waveform_from_pcs} using all 15 measured principal component parameters. We then determine the postbounce oscillation frequency using the the approximate posterior waveform. Evaluating $\fpeak$ for all the samples gives a posterior probability distribution for $\fpeak$. Comparing the posterior with Table~\ref{tab:average_f_peak_values} enables us to rule out the equations of state inconsistent with the signal waveform. In this way gravitational waves from core-collapse provide us a different regime than binary neutron star mergers to study the nuclear equation of state.  

\section{Parameter Estimation} \label{section:pe}

By combining the methods described above with Bayesian parameter estimation~\cite{Bayes:1763,Jaynes:2003jaq} we can estimate the posterior probability distributions for the physical parameters of astrophysical signals. Our Bayesian parameter estimation samples the probability of the modeled parameter values given a model and set of detectors' data using Markov Chain Monte Carlo methods. We calculate the posterior probability density function, $\posterior$, for the set of parameters $\vec{\pset}$ for the gravitational-waveform model, $H$, given the gravitational-wave data from the detectors $\vec{d}(t)$
\begin{equation}
\label{eqn:bayes} \posterior =
\frac{\likelihood \prior}{\evidence} ,
\end{equation}
where $\prior$ is the prior---the assumed knowledge of the distributions for the parameters $\vec{\pset}$ describing the signal, before considering the data. $\likelihood$ is the likelihood---the probability of obtaining the data $\vec{d}(t)$ given the model $H$ with parameters 
$\vec{\pset}$. We use the Gaussian likelihood in this analysis, which is given by \cite{1970esn..book.....W}:
\begin{equation*}
    \likelihood = \exp \left[ - \frac{1}{2} \sum_{i=1}^{N} \braket{\tilde{n}_i (f)| \tilde{n}_i(f)}\right] 
\end{equation*}
\begin{equation}
= \exp \left[  - \frac{1}{2} \sum_{i=1}^{N} \braket{\tilde{d}_i (f) - \tilde{s}(f, \pset)| \tilde{d}_i(f)  - \tilde{s}(f, \pset)}\right]
\end{equation}
where $N$ is the number of detectors (in our case, $N = 1$), and  $\tilde{d}_i(f)$ and $\tilde{n}_i(f)$ are the Fourier transforms of the data and the noise in the detector. We sample the posterior probability distribution using stochastic sampling methods. Our choice of sampler in \texttt{PyCBC Inference} \cite{Biwer:2018osg} is guided by the fact that the default parallel tempered MCMC sampler \texttt{emcee}\_\texttt{pt} \cite{ForemanMackey:2012ig, vousden:2016url, vousden2016dynamic} can experience problems converging for signals with signal-to-noise ratios greater than 100.  To address this, we use the dynamic nested sampling package \texttt{Dynesty} \cite{2020MNRAS.493.3132S, 2004AIPC..735..395S, Skilling:2006gxv} which provides posterior probability distributions for all the signals explored here. For signals with very high signal-to-noise ratio, the detector noise becomes negligible and so it is possible to obtain a point estimate of the signal parameters by directly computing the inner product between the signal and the basis vectors. By performing this spot-check for the high signal-to-noise ratio signals, we find that these point estimates agree with the posteriors obtained by the \texttt{Dynesty} sampler.
 
In our analysis, we assume that any gravitational-wave signal from a core-collapse supernova will be accompanied by a neutrino signal detected by neutrino observatories such as IceCube \cite{Abbasi:2011ss}, Super-Kamiokande \cite{Ikeda:2007sa} or DUNE \cite{Acciarri:2015uup}. The neutrino observations can estimate the time of the core bounce to within $3-4$~ms \cite{Yokozawa:2014tca, Muller:2019upo, Tamborra:2017ull}. Our analysis only considers the core bounce and the next $5-7$~ms, and we use assume that information from the neutrino observations can provide a narrow prior of $8$~ms for the time of the bounce. We also assume that the distance and sky location to the source are known and we do not include them in the parameter estimation.

We use \texttt{PyCBC Inference} \cite{Biwer:2018osg} to obtain posteriors for the coefficients of the first fifteen principal components of the waveform catalog. We use uniform priors for all the fifteen coefficients as shown in Table~\ref{tab:priors}, in addition to the constraint that the samples are restricted with the convex hull formed by the point cloud of the first three model parameters for the waveforms in the catalog. Using the map discussed in section \ref{section:map} and the methods to extract $\fpeak$ values, we translate the posteriors on the coefficients to posteriors on $\beta$ and $\fpeak$. 
\begin{table}
\begin{center}
\begin{tabular}{c c c} 
\hline
\hline
Parameter & Lower bound on prior  & Upper bound on prior \\ 
\hline   
$\alpha_1$        & 0.0 & 10.5  \\ 
$\alpha_2$        & -5.0 & 3.55  \\
$\alpha_3$        & -2.0 & 2.0  \\ 
$\alpha_4$        & -1.5 & 2.0  \\ 
$\alpha_5$        & -1.0 & 1.75  \\
$\alpha_6$        & -0.85 & 1.05  \\
$\alpha_7$        & -0.75 & 1.5  \\
$\alpha_8$        & -0.75 & 0.75  \\
$\alpha_9$        & -0.75 & 0.75  \\
$\alpha_{10}$        & -0.75 & 0.75  \\
$\alpha_{11}$        & -0.75 & 0.75  \\
$\alpha_{12}$        & -0.75 & 0.75  \\
$\alpha_{13}$        & -0.75 & 0.75  \\
$\alpha_{14}$        & -0.75 & 0.75  \\
$\alpha_{15}$        & -0.75 & 0.75  \\
$t_{\mathrm{bounce}} (\mathrm{GPS~time})$        &1126259469.517 & 1126259469.525  \\ 
\hline
\end{tabular}
\caption{Upper and lower bounds on the uniform priors used for the model parameters $\alpha_i$ and $t_{\mathrm{bounce}}$ in Bayesian parameter estimation. The values for $\alpha_i$ were chosen based on the range of values obtained from the construction of principal component basis set. $t_{\mathrm{bounce}}$ has a uniform prior width of $8$ms. All signals are aligned such that the bounce is at $t_{\mathrm{GPS}} = 1126259469.5 + 0.02125$ where $0.02125$ is the light travel time between the center of the Earth and the detectors. Note that an additional constraint on the priors is to restrict the samples with the convex hull formed by the first three model parameters of the waveforms in the catalog (see Sec.~\ref{section:map}).} \label{tab:priors}
\end{center}
\end{table}

\section{Results}\label{section:results}

\begin{figure}[t]
  \includegraphics[width=\columnwidth]{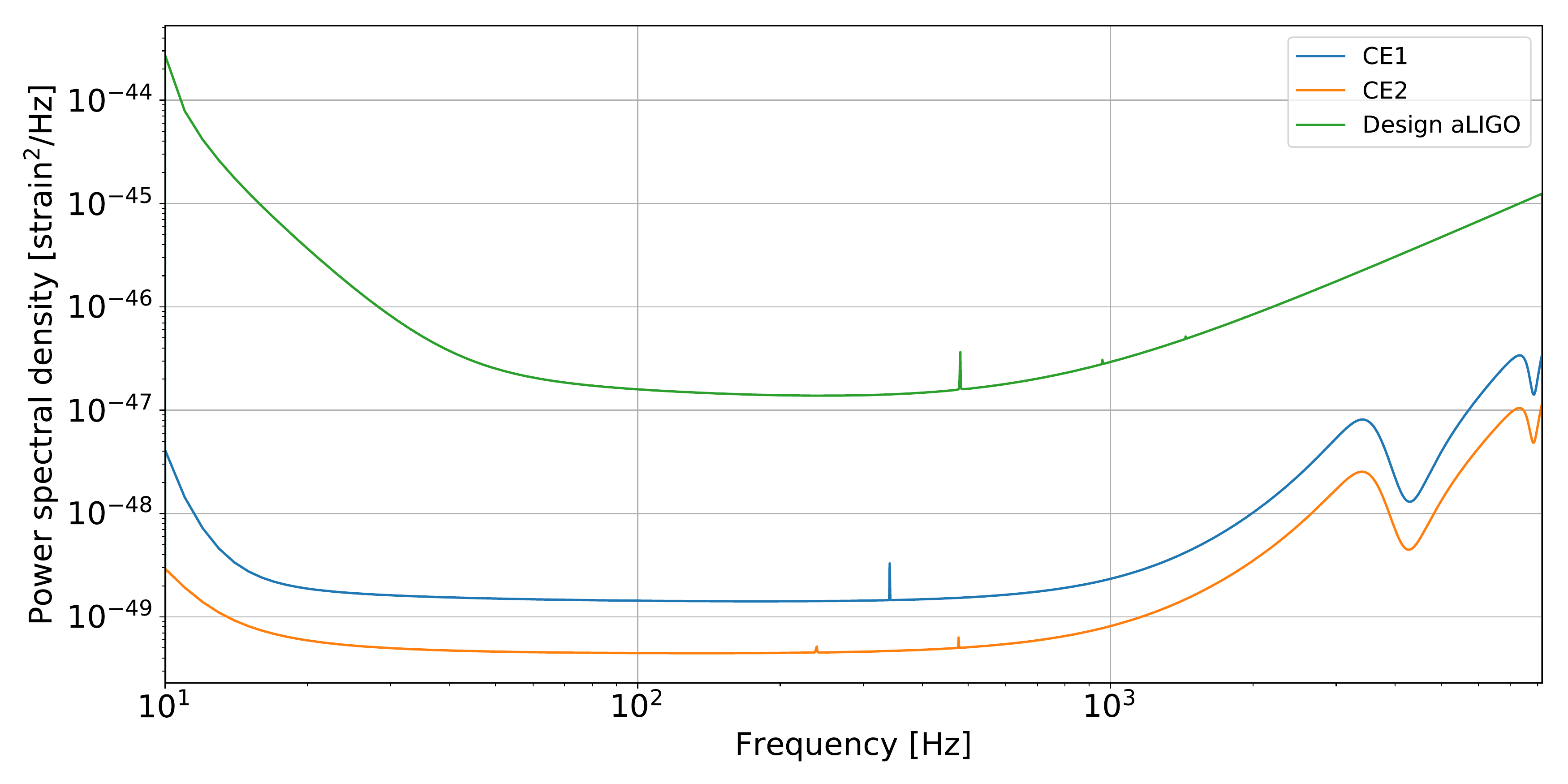}
  \vspace*{-0.5cm}%
  \caption{Predicted noise power spectral densities for Advanced LIGO, Cosmic~Explorer~1, and Cosmic Explorer 2 detectors.}
\label{fig:noise_curves}
\end{figure}

\begin{figure}[t]
  \includegraphics[width=\columnwidth]{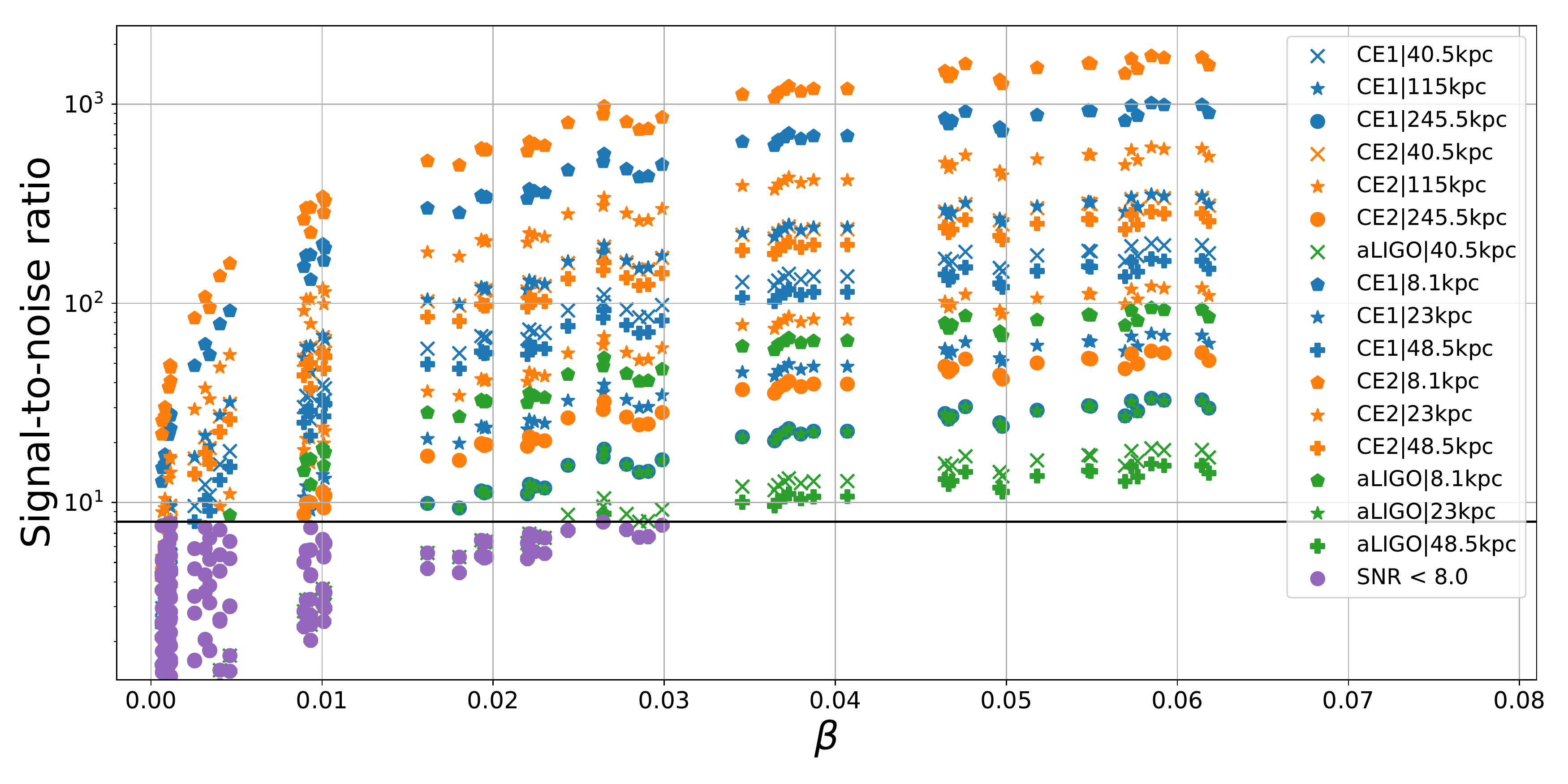}
  \vspace*{-0.5cm}
  \caption{The vertical axis shows the signal-to-noise ratios of waveforms used as astrophysical signals. The horizontal axis shows the $\beta$ of the core progenitor at bounce. These sources are assumed to be at distances of $8.1$~kpc, $23$~kpc, $40.5$~kpc, $48.5$~kpc, $115$~kpc, and $242$~kpc and the signals are observed in the Cosmic Explorer 1 (CE1), Cosmic Explorer 2 (CE2), and Advanced LIGO (aLIGO) gravitational wave detectors. We ignore the waveforms with signal-to-noise ratios below 8 (shown as purple dots) and do not perform parameter estimation on them.}
\label{fig:beta_SNR}%
\vspace*{-0.5cm}%
\end{figure}

We test our method using the 60 signal waveforms reserved from above. Each waveform, consisting of the core-collapse, postbounce oscillation, and prompt convection phases, is used to create a simulated observation by adding it to Gaussian noise colored to the strain sensitivity of the Advanced LIGO detectors and the third-generation detectors: Cosmic Explorer 1 (CE1), and Cosmic Explorer 2 (CE2). Cosmic Explorer is the proposed third generation detector which is planned to begin observing in 2030s \cite{Reitze:2019iox}. The first stage of the observatory, Cosmic Explorer 1, is the scaling up of the Advanced LIGO technologies to an interferometer with 40~km arm length. The second stage of the observatory, Cosmic Explorer 2, will be an upgrade on the core optics of Cosmic Explorer 1 by using cryogenic technologies and new mirror substrates. The predicted noise power spectral densities of the three detectors used in this study are shown in Fig.~\ref{fig:noise_curves}. We place the sources at distances corresponding to the center of the Milky Way galaxy (8~kpc), far edge of the Milky Way from the Earth (23~kpc), the Large Magellanic Cloud (48.5~kpc), and out to $242$~kpc to capture the dwarf satellite galaxies of the Milky Way in the local group. In addition, we place the sources at the distances of $40.5$~kpc and at $115$~kpc. The sources are assumed to be optimally oriented for the detector. The signal-to-noise ratio of the signal waveforms and its variation with $\beta$ is plotted in Fig.~\ref{fig:beta_SNR}. We do not perform the analysis if the simulated signal has a signal-to-noise ratio less than 8 (shown as purple points in the figure). We note that more sensitive interferometers are able to detect more number of signals with low $\beta$. Advanced LIGO is not able to detect any sources at $115$~kpc or beyond. It is also unable to detect the sources with $\beta < 0.02$ at $40.5$~kpc and beyond. The signal-to-noise ratios and detection ranges in our study are consistent with those obtained for comparable signals in previous core-collapse supernovae search studies \cite{Abbott:2019pxc, Gossan:2015xda}.

We summarize our results in Table~\ref{tab:results}. We measure the median values and the $90\%$ credible intervals from the posteriors obtained from MCMC for $\beta$ and $\fpeak$. The width of $90\%$ credible intervals show how precisely we can measure the parameters. $90\%$ credible interval of $\fpeak$ is useful to determine the equations of state consistent with the signal, using Table~\ref{tab:average_f_peak_values}. The mean of the median values provides an estimate of the accuracy of the measurement of the parameters.  We present our results by classifying the signals in two sets: $\beta < 0.02$, and $ \beta \geq 0.02$. 
\begin{table*}
\centering
\begin{tabular}{|c|c|c|c|c|c|c|c|}
\hline
\hline
\multirow{2}{*}{Detector}& \multirow{2}{*}{\shortstack[c]{Source \\ distance [kpc]}} & \multirow{2}{*}{$\beta$ range}& \multirow{2}{*}{\shortstack[c]{\\ Number \\ of \\ signals}} & \multicolumn{2}{c|}{$\beta$} & \multicolumn{2}{c|}{$\fpeak$} \\
\cline{5-8}
 &   &  &   & \makecell{Mean $90\%$ \\ credible interval} & \shortstack[c]{Mean \\ fractional\\ error}  & \shortstack[c]{Mean $90\%$ \\ credible \\ interval [Hz]} &\shortstack[c]{Mean \\ fractional\\ error} \\
\hline
\multirow{6}{*}{Advanced LIGO} & \multirow{2}{*}{8}   & $\beta < 0.02$ & 13 &  0.004 & 22 $\%$ & 289 & 26 $\%$  \\
                               \cline{3-8}
                               &                        & $\beta \geq 0.02$ & 35 & 0.004 & 10 $\%$ & 7 & 4 $\%$  \\
                               \cline{2-8}
                               & \multirow{2}{*}{23}   & $\beta < 0.02$ & 5 &  0.01 & 19 $\%$ & 780 & 3 $\%$  \\
                               \cline{3-8}
                               &                        & $\beta \geq 0.02$ & 35 & 0.009 & 13 $\%$ & 39 & 4 $\%$  \\
                               \cline{2-8}
                               & \multirow{2}{*}{48.5}   & $\beta < 0.02$ & 0 & -- & -- & -- & --  \\
                               \cline{3-8}
                               &                        & $\beta \geq 0.02$ & 25 & 0.02 & 12 $\%$ & 57 & 3 $\%$  \\
\hline
\multirow{8}{*}{Cosmic Explorer 1} & \multirow{2}{*}{8}  & $\beta < 0.02$ & 25 & 0.0004 & 26 $\%$ & 37 & 18 $\%$  \\
                                        \cline{3-8}
                                    &                     & $\beta \geq 0.02$ & 35 & 0.0008 & 6 $\%$ &  2& 3 $\%$  \\
                                        \cline{2-8}
                                    & \multirow{2}{*}{23}  & $\beta < 0.02$ & 20 & 0.001 & 21 $\%$ & 147 & 11 $\%$  \\
                                        \cline{3-8}
                                    &                     & $\beta \geq 0.02$ & 35 & 0.002 & 6 $\%$ &  5& 3 $\%$  \\
                                        \cline{2-8}
                                    & \multirow{2}{*}{48.5}  & $\beta < 0.02$ & 17 & 0.002 & 15 $\%$ & 167 & 5 $\%$  \\
                                        \cline{3-8}
                                    &                     & $\beta \geq 0.02$ & 35 & 0.003 & 7 $\%$ &  11& 3 $\%$  \\
                                        \cline{2-8}
                                    & \multirow{2}{*}{242}  & $\beta < 0.02$ & 5 & 0.007 & 6 $\%$ & 205 & 2 $\%$  \\
                                        \cline{3-8}
                                    &                     & $\beta \geq 0.02$ & 35 & 0.009 & 8 $\%$ &  64& 2 $\%$  \\
                                        \cline{2-8}
\hline
\multirow{8}{*}{Cosmic Explorer 2} & \multirow{2}{*}{8}  & $\beta < 0.02$ & 25 & 0.0002 & 27 $\%$ & 4 & 21 $\%$  \\
                                        \cline{3-8}
                                    &                     & $\beta \geq 0.02$ & 35 & 0.0005 & 7 $\%$ &  1& 3 $\%$  \\
                                        \cline{2-8}
                                    & \multirow{2}{*}{23}  & $\beta < 0.02$ & 24 & 0.0005 & 24 $\%$ & 120 & 14 $\%$  \\
                                        \cline{3-8}
                                    &                     & $\beta \geq 0.02$ & 35 & 0.001 & 7 $\%$ &  3& 3 $\%$  \\
                                        \cline{2-8}
                                    & \multirow{2}{*}{48.5}  & $\beta < 0.02$ & 18 & 0.001 & 18 $\%$ & 22 & 10 $\%$  \\
                                        \cline{3-8}
                                    &                     & $\beta \geq 0.02$ & 35 & 0.002 & 7 $\%$ &  6& 3 $\%$  \\
                                        \cline{2-8}
                                    & \multirow{2}{*}{242}  & $\beta < 0.02$ & 11 & 0.004 & 8 $\%$ & 227 & 3 $\%$  \\
                                        \cline{3-8}
                                    &                     & $\beta \geq 0.02$ & 35 & 0.006 & 8 $\%$ &  51& 3 $\%$  \\
                                        \cline{2-8}

\hline
\end{tabular}
\caption{The table summarizes the results of parameters estimation of $\beta$ and $\fpeak$ for signal sources at $8$~kpc, $23$~kpc, $48.5$~kpc, and $242$~kpc. We have categorized the results for the all the signals on basis of the detector they are observed in, their distance and the corresponding value of $\beta$. We present the mean of the $90\%$ credible interval widths and the mean value of the errors and from the posteriors obtained for $\fpeak$ and $\beta$. The average $90\%$ credible interval width of $\beta$ for sources at $8$~kpc observed in Advanced LIGO is $0.004$, while for the third generation detectors its an order of magnitude less. The precision to which $\beta$ can be measured decreases when the $\beta$ of the signal waveform increases, or the source distance increases. Note that the method to measure $\fpeak $ for signals with $\beta < 0.02$ is unreliable. We include these results here for completeness.} \label{tab:results}
\end{table*}

\begin{figure}[t]
  \includegraphics[width=\columnwidth]{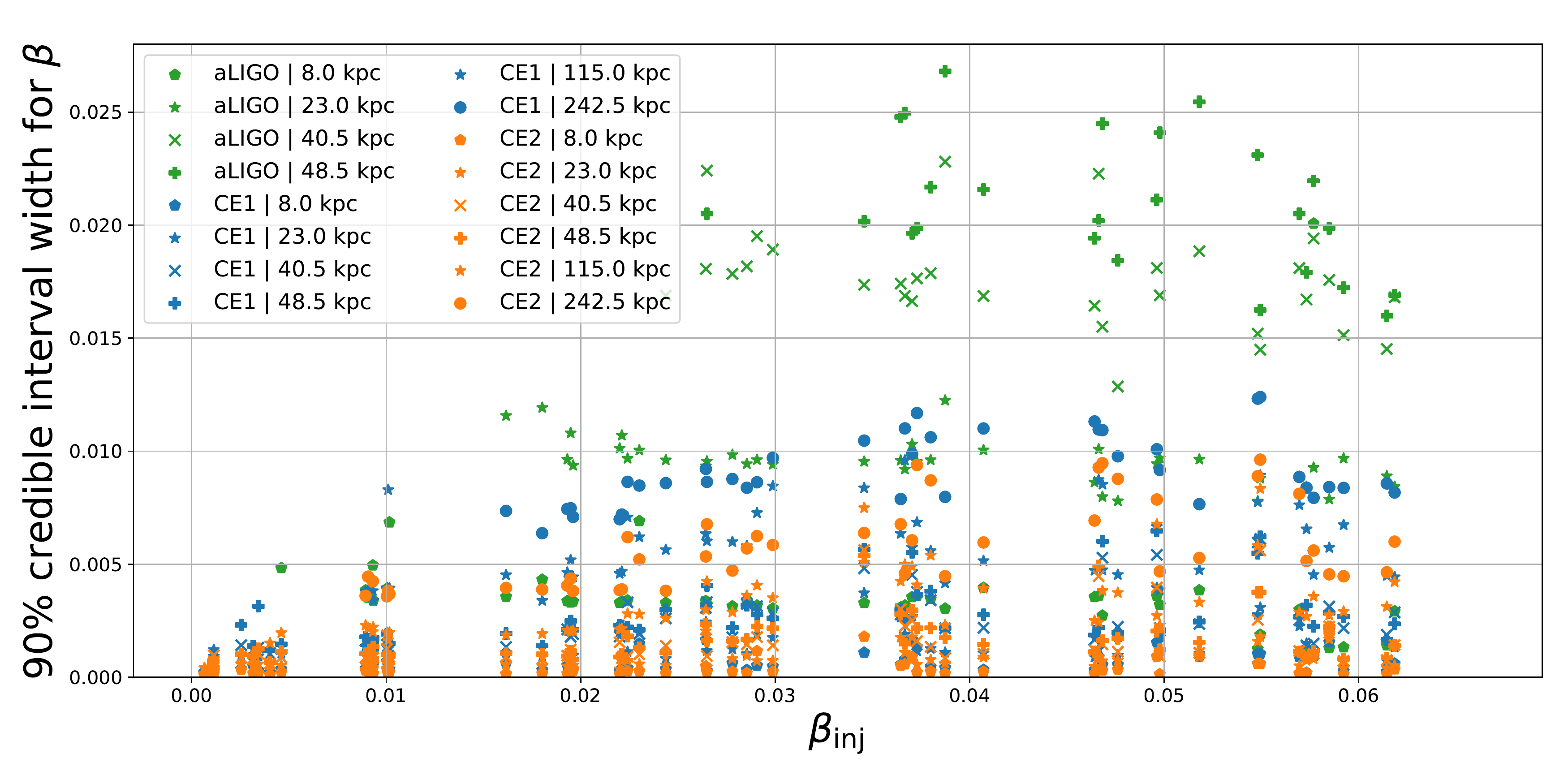}
  \vspace*{-0.5cm}
  \caption{The $90\%$ credible interval width of the posteriors obtained for $\beta$ as a function of the $\beta$ of the injected signal waveform. We note that the signals observed in Cosmic Explorer 1 (blue) and Cosmic Explorer 2 (orange) are measured an order of magnitude more precisely than the signals in Advanced LIGO (shown in green). On an average, the $90\%$ credible interval width for signals observed in Cosmic Explorer 1 is $1.5$ times that of the signals observed in Cosmic Explorer 2.} 
\label{fig:cred_int_width_vs_injected_beta}%
\vspace*{-0.5cm}%
\end{figure}

The mean width of the $90\%$ credible interval for $\beta$ for signals sources at the center of the Milky Way with $\beta=0.04$ is $0.004$ when observed in Advanced LIGO, improving to a width of $0.0008$ if observed in Cosmic Explorer detectors. For sources at $48.5$ kpc it increases to $0.02$ for Advanced LIGO detections and $0.003$ for Cosmic Explorer detections. We note that the width of the $90\%$ credible intervals increases as the source distance increases. In addition to that, as the value of $\beta$ of the injected signal increases the $90\%$ credible interval width also increase, even though the signal-to-noise ratio also increases. As discussed in Sec.~\ref{section:map}, this is because the coefficients for known values of $\beta$ used to construct the map become sparse for higher values of $\beta$ and the interpolation suffers. On an average, the $90\%$ credible interval width for signals observed in Cosmic Explorer 1 is $1.5$ times that of the signals observed in Cosmic Explorer 2. Fig.~\ref{fig:cred_int_width_vs_injected_beta} shows the $90\%$ credible interval width of the posteriors of $\beta$ as a function of the injected value of $\beta$ for all the signals. For the sources at a given distance observed in a particular detector, the $90\%$ credible interval does not vary significantly across the range of injected values of $\beta$. For some signals with $\beta <0.02$, the signal-to-noise ratio is less than 8, and hence we do not perform parameter estimation on them.

\begin{figure}[t]
  \includegraphics[width=\columnwidth]{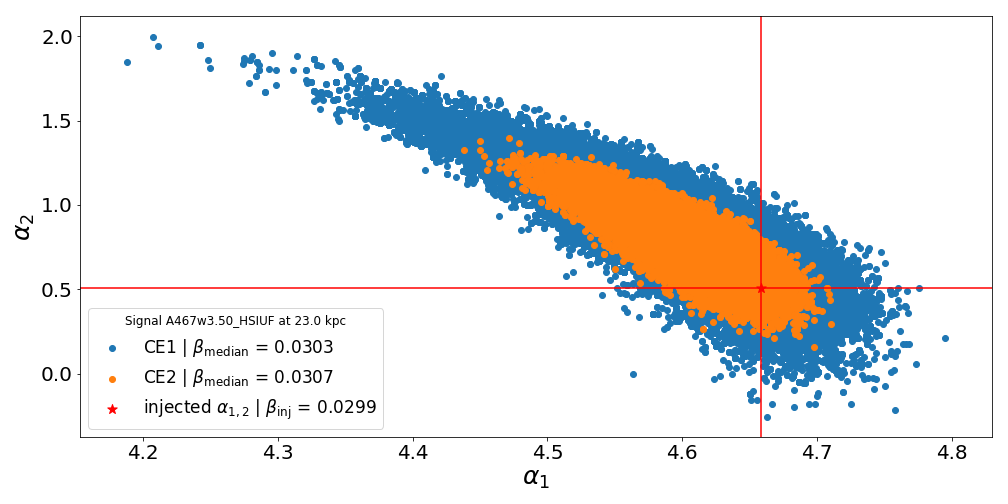}
  \vspace*{-0.5cm}
  \caption{The $\alpha_1$ and $\alpha_2$ posteriors obtained for the signal with $\beta = 0.0299$ at $23$~kpc observed in Cosmic Explorer 1 (shown in blue) and Cosmic Explorer 2 (shown in orange). The $(\alpha_1, \alpha_2)$ point corresponding to the injected signal (shown as the red star) is within the $90\%$ contour region of both posteriors. The $90\%$ contour region for the posterior of signal observed in Cosmic Explorer 2 is smaller than that of Cosmic Explorer1 because the signal has higher signal-to-noise ratio in the former. However, when these posteriors are transformed into the posteriors of $\beta$, the error in median values of $\beta$ is larger for Cosmic Explorer 2 than Cosmic Explorer 1.} 
\label{fig:posterior}%
\vspace*{-0.5cm}%
\end{figure}

For signals sources at a distance of $23$~kpc with $\beta <0.02$ observed in Cosmic Explorer 1, we estimate $\beta$ with an error of $21\%$. This increases to $24\%$ for Cosmic Explorer 2. For signal sources at $23$~kpc with $\beta > 0.02$, we can estimate $\beta$ with $6\%$ error for Cosmic Explorer detectors. The error increases as the source distance increases. Fig.~\ref{fig:posterior} shows the $\alpha_1$ and $\alpha_2$ posteriors obtained for the signal with $\beta = 0.0299$ at a distance of $23$~kpc observed in Cosmic Explorer 1 (blue) and Cosmic Explorer 2 (orange). Since the signal is observed with higher signal-to-noise ratio in Cosmic Explorer 2 than in Cosmic Explorer 1, the posteriors obtained for the former are smaller in area. However, the point with $\alpha_1$ and  $\alpha_2$ values corresponding to the signal (shown as the red star) is within the $90\%$ credible region of both posteriors. When these posteriors are translated to the posteriors of $\beta$, using the map discussed in Sec.~\ref{section:map}, the difference between the median value of $\beta$ obtained and the $\beta$ of the injected signal is higher for Cosmic Explorer 2 than that for Cosmic Explorer 1. Such error is introduced for several signals and leads to lower overall error for Cosmic Explorer 1 than its upgraded counterpart. For Advanced LIGO, $\beta$ is measured with an error of $9\%$. 

For signals with $\beta \geq 0.02$ observed in the third generation detectors, we can measure $\fpeak$ with an mean error of upto $3\%$. The average $90\%$ credible intervals obtained for $\fpeak$ for such signals within the galaxy is $5$~Hz. Estimating $\fpeak$ with such precision restricts the possible equations of state consistent with the $\fpeak$ values, specially for signals with $ 0.02 \leq \beta \leq 0.06$. We obtain an average $90\%$ credible intervals for $\fpeak$ of $7$ Hz for signals at the center of Milky way observed in Advanced LIGO noise, with a systematic error of $4\%$. For sources that are further away, the average $90\%$ credible interval are more that $35$~Hz. The systematic error is larger that the range spanned by the mean $\fpeak$ values of various equations of state listed in Table~\ref{tab:average_f_peak_values} and we conclude that third-generation gravitational-wave detectors are required to extract nuclear physics from core-collapse supernovae. The method to extract $\fpeak$ for any waveform with a corresponding $\beta \leq 0.02$ is unreliable, and hence we get large systematic errors and $90\%$ credible intervals for such signals. We include these results for completeness.

\section{Conclusion} \label{section:conclusion}
Practical implementation of Bayesian inference relies on the existence of parameterised gravitational-waveform models that are inexpensive to compute. Such models, with parametrization for the core rotation rate and the postbounce oscillation frequency, do not exist for complete core-collapse supernovae waveforms due to the complexity of the physics involved. In this paper, we address this problem for the first two phases of core-collapse signals, namely the core bounce and the postbounce oscillations. We use principal component analysis to create a parameterised model that extracts the most common features of the bounce signal onto the principal components. We construct a map between the physical parameters and the model parameters (principal components and their coefficients). We use Bayesian inference to measure the coefficients of the first fifteen principal components for a signal observed in gravitational-wave detectors, and use the inverse of the aforementioned map to obtain posteriors of the physical parameters. In particular, we obtain posterior probability distributions for the ratio of rotational kinetic energy to the potential energy of the core at bounce ($\beta$) and the peak frequency of the post bounce oscillations of the protoneutron star ($\fpeak$). 

$\beta$ depicts the rotation rate of the inner core of the star at the core bounce. We find the relationship between the model parameters and $\beta$ by interpolating known values of $\beta$ from the hyper-volume formed by the model parameters. $\fpeak$ encodes useful information about the nuclear equation of state, and tells us about the behaviour of hot, dense nuclear matter in the core of the star. We can successfully measure $\fpeak$ for waveforms with $\beta \geq 0.02$, however the method to extract it fails for waveforms of extremely slowly rotating cores. 

For signals with $\beta \geq 0.02$ at a distance of $8$ kpc detected in Advanced LIGO, $\beta$ can be estimated with a $90\%$ credible interval of $0.004$ for Advanced LIGO, and $0.0008$ for Cosmic Explorer detectors. The width of the $90\%$ credible interval for $\beta$ increases to $0.002$ ($0.003$) for sources at $23$~kpc ($48.5$~kpc). On an average, the $90\%$ credible interval for $\beta$ for signals observed in Cosmic Explorer 1 is $1.5$ times larger than that for signals observed in Cosmic Explorer 2.  We can also estimate $\fpeak$ to within $\sim 6$ Hz for signals sources upto the distance of $48.5$~kpc with $\beta \geq 0.02$ observed in the third-generation detectors. Using the posteriors on $\fpeak$, we can successfully rule out the nuclear equations of state that are inconsistent with the signal. The error in measuring $\fpeak$ for the signals observed in Advanced LIGO is $4\%$ with an average $90\%$ credible interval width of $6$~Hz for sources at the center of the Milky Way. For sources that are further away, the $90\%$ credible interval width increases to more than $20$~Hz. We conclude that third-generation detectors are required to constrain the nuclear equation of state from gravitational-wave observations of core-collapse supernovae.

Previous studies have used principal component analysis in Bayesian reconstruction of the signal observed in the detectors \cite{Rover:2009ia, Suvorova:2019ebd} or to infer the core-collapse explosion mechanism \cite{Logue:2012zw, Powell:2016wke, Powell:2017gbj, Roma:2019kcd}. Edwards et al \cite{Edwards:2014uya} used principal component analysis to measure $\beta$ for signals observed in Advanced LIGO with signal-to-noise ratio 20, and obtained the 90\% confidence interval width of 0.06. We demonstrate a method that uses principal component analysis in Bayesian estimation of physical parameters $\beta$ and $\fpeak$, and to find the dependence of gravitational-waveform morphology on these physical parameters. For a signal comparable to the ones in Edwards \textit{et al.} study, our method yields a confidence interval of 0.02, which is three times smaller than that found by Edwards \textit{et al.}

A more robust map between the model parameters and $\beta$ can be constructed my populating the model parameter space and using machine learning. We leave the construction of this map and analysis of signals observed in Einstein telescope for future work. 

\section*{Acknowledgments}
We thank Adam Burrows, Daniel Finstad, and Chris Fryer for helpful discussions. The authors were supported by National Science Foundation awards PHY-1707954, PHY-1836702, and PHY-2011655. The authors thank the Kavli Institute for Theoretical Physics for hospitality and partial support from National Science Foundation award PHY-1748958. Computational work was supported by Syracuse University and National Science Foundation award OAC-1541396. Supporting data for this manuscript is available from \url{https://github.com/sugwg/sn-core-bounce-pe}
\bibliography{main}

\end{document}